\documentclass[review]{elsarticle}

\usepackage{lineno,hyperref}
\usepackage[pages=some]{background}
\usepackage{amsmath}
\usepackage{amssymb}
\usepackage{rotating}
\usepackage[super]{nth}
\usepackage[framed,numbered]{matlab-prettifier}
\usepackage{epstopdf}
\usepackage{algorithm}
\usepackage[noend]{algpseudocode}
\usepackage{booktabs}
\usepackage{multirow}
\usepackage{array} 
\usepackage{colortbl}
\usepackage{array}
\usepackage{titlecaps}
\usepackage{framed} 
\usepackage{enumitem}
\usepackage{textcomp}
\usepackage[algo2e]{algorithm2e}
\usepackage{adjustbox}
\definecolor{MatlabCellColour}{RGB}{252,251,220}
\usepackage{listings}
\modulolinenumbers[5]

\begin{document}
\date{}
\begin{frontmatter}

\title{Global Outliers Detection in Wireless Sensor Networks: A Novel Approach Integrating Time-Series Analysis, Entropy, and Random Forest-based Classification}

 \author{Mahmood~Safaei$^{1}$, Maha Driss$^{2}$, Wadii Boulila$^{2}$, Elankovan A Sundararajan$^{3}$, and Mitra~Safaei$^{4}$}

\address{%

$^{1}$ \quad 6G Innovation Centre, University of Surrey, UK.\\
$^{2}$ \quad RIADI Laboratory, University of Manouba, Tunisia. \\
$^{3}$ \quad Center for Software Technology and Management, Faculty of Information Science and Technology, University Kebangsaan Malaysia, Malaysia. \\
$^{4}$ \quad Fakultät Electronic und Informatik, Gottfried Wilhelm Leibniz Universität Hannover, Germany.\\
}



%
%

\begin{abstract}
Wireless Sensor Networks (WSNs) have recently attracted greater attention worldwide due to their practicality in monitoring, communicating, and reporting specific physical phenomena. The data collected by WSNs is often inaccurate as a result of unavoidable environmental factors, which may include noise, signal weakness, or intrusion attacks depending on the specific situation. Sending high-noise data has negative effects not just on data accuracy and network reliability, but also regarding the decision-making processes in the base station. Anomaly detection, or outlier detection, is the process of detecting noisy data amidst the contexts thus described. The literature contains relatively few noise detection techniques in the context of WSNs, particularly for outlier-detection algorithms applying time series analysis, which considers the effective neighbors to ensure a global-collaborative detection. Hence, the research presented in this paper is intended to design and implement a global outlier-detection approach, which allows us to find and select appropriate neighbors to ensure an adaptive collaborative detection based on time-series analysis and entropy techniques. The proposed approach applies a random forest algorithm for identifying the best results. To measure the effectiveness and efficiency of the proposed approach, a comprehensive and real scenario provided by the Intel Berkeley Research lab has been simulated. Noisy data have been injected into the collected data randomly. The results obtained from the experiment then conducted experimentation demonstrate that our approach can detect anomalies with up to 99\% accuracy.

\end{abstract}

\begin{keyword}
Wireless sensor network; anomaly detection; outlier detection; time series analysis; entropy; Random Forest.
\end{keyword}

\end{frontmatter}

\section{Introduction}
Wireless sensor networks (WSNs) are drawing great interest worldwide, especially with the considerable progress of technologies that are leading to the apparition and enhancement of small smart sensors. With their reduced size, limited computing units, and condensed processing resources, these sensors are cheaper than their traditional counterparts. The nodes embedded in smart sensors enable them to detect data, measure it, and collect it from various points in the target environment. In addition, these nodes transfer sensory data into the sink, or base station, of the sensor, where decisions are processed and made. These capacities mean that smart sensor nodes have low power requirements and are relatively simple devices despite their complex functions: most consists of the nodes themselves plus a power supply, processor, radio transmitter,  memory, and actuator \cite{Yick2008}.\\
A WSN is composed of multiple such wireless sensor devices, sometimes hundreds or thousands, implemented in a location determined by the user \cite{Akyildiz2002}. With WSNs, reliable communication is very important, and the literature has proposed several algorithms intended to guarantee a WSN's reception of reliable, less noisy data. Outlier detection algorithms have been listed in parts of the literature, but they have not been studied in as much depth as some other options.\\
An outlier is defined by \cite{Hawkins1980} as "an observation that diverges to a large extent from other observations to give rise to doubts that it was produced by a separate method". In \cite{Ord1996}, an outlier represents "an observation (or a set of observations) that seems to be inconsistent with the rest of the data in that set". Another definition of outliers as they relate to WSNs has also been provided by \cite{Kandhari2009}, which is "the measurements that show significant deviation from the typical pattern of sensed data".\\
There are several sources of outliers, which are detected in the data collected by WSNs such as event detection \cite{Ding2005, Chen2005, Martincic2006, zhang2012statistics}, fault detection \cite{Chen2006, Luo2006}, and intrusion detection \cite{DaSilva2005,Bhuse2006}.\\
In general, outliers can be classified into two different categories, local or global \cite{ayadi2017outlier}. The category that any particular outlier falls into can be determined based on the types and range of data surveyed and utilized in the process of detecting it \cite{Subramaniam2006}. The detection of local outliers is performed by considering a single sensor node and carried out either by identifying irregular values at the considered node on the basis of its own values collected previously or by using data from that node's neighbors. The outlier detection process in the second approach offers greater accuracy than the processes of the first; since this second approach takes into account the benefits that are gained from spatio-temporal correlations among the overall collected sensor data \cite{gupta2013outlier,safaeistandalone}. In addition, the second approach detects outliers in a more global perspective, which it accomplishes by considering the whole network. This also makes it possible to detect global noisy data at distinct network levels by considering the network typology \cite{Meratnia2010}. In the case of centralized network architecture, all data are collected in the main sink node, which is where the outliers’ detection is also performed. The main drawback of this latter method is the way in which it both increases overall response time and also generates additional costs for communication \cite{Subramaniam2006}.\\
In much of the related literature, several other methods have also been proposed for implementing outlier detection, which have included statistical modeling, information theory, Z-Score, and data mining-based methods \cite{Kandhari2009}. The data mining-based method denotes the discovery of valuable and interesting information from extensive sources of data, and in this context, outlier detection in WSNs would be an appropriate area of application of this method \cite{Pang-Ning2006} \cite{Han2006}.\\
In recent years, the possibility for a quick, efficient, and accurate means of detecting outliers in WSNs has become of great interest to researchers since it can guarantee robust functionality of the affected network, the reliability of data thus collected and analyzed, and the generation of real-time event reports \cite{safaei2020systematic}. In addition, the detection of outliers in WSNs guarantees the analysis of the validity of the data and therefore reduces the communication costs of incorrect data. Furthermore, potential attacks on the network can be identified through the detection of outliers, which in turn can lead to an improvement of the network security.\\
In this paper, we suggest a new approach to outlier detection, one with its basis in time-series modeling and forecasting with neighbors’ collaboration. First, we start by extracting features allowing the time-series modeling and forecasting. Then, an adaptive entropy-based method is proposed to determine neighbor spatial- correlation. The third step of the proposed approach aims to determine the outliers and the anomaly data in each sensor node by performing a random forest classification algorithm.\\ 
The main contributions of the present work can be summarized in the following 3 points:
\begin{itemize}
    \item The formulation of the problem of outlier detection in WSNs as a time-series analysis problem by considering the historically collected data;
    \item The proposition of an entropy-based method to select the best neighbor related to a considered sensor in order to ensure a spatio-temporal correlation useful for the outlier detection. The current work focuses on evaluating the importance of temporal features and the correlation among the data of time-series data for the detection of outliers. 
The spatio-temporal correlation can be exploited to improve the overall network performance. The characteristics of the correlation in the WSN context can be classified into spatial and temporal correlations \cite{vuran2004spatio}. The first one relies on multiple sensors recording the same event. In this case, data are highly correlated with the recorded observations. For the second case, temporal correlations are recorded for many WSNs applications such as event tracking or area monitoring, especially when nodes periodically transmit observations about event features. Moreover, spatio-temporal correlation can bring important advantages when developing efficient communication protocols for the considered WSNs. For instance, data coming from spatially separated sensors are more important to the sink than highly correlated data from nodes in proximity \cite{vuran2004spatio}. Additionally, in the case of event tracking, temporal correlations play an important role in adjusting the frequency of measurement reporting which is essential in order to minimize energy expenditure. To the best of our knowledge, numerous research studies have been conducted about the outlier detection problem in WSNs but most of them mainly detect anomalies using offline data and few studies detect outliers using stream data. Offline anomaly detection can affect real-time decision-making, which conflicts with the WSN reliability concept. In addition, traditional outlier detection methods such as those based on a fixed threshold are not efficient since space and temporal conditions are changing dynamically. Therefore, reading data from neighbor nodes for spatial data will increase the accuracy of the proposed algorithm;
       \item  The development and application of a random forest-based algorithm using time-series data to globally identify outliers in each sensor node. This algorithm prevents from making incorrect decisions on the base station and also increases the lifetime of the network.
\end{itemize}

The remainder of this work is structured in the following way: in Section 2, the relevant literature and research on outlier and anomaly detection in WSNs are reviewed. In Section 3, the approach we propose is described in greater detail. In Section 4, experimental results carried out on a synthetic and real-world dataset (provided by Intel Berkeley Research lab) are reported and analyzed. Section 5 features concluding remarks on our results and consideration of future directions for related work.

\section{Related works}

Detecting outliers in WSNs is a challenging problem due to certain characteristics of sensors: resource constraints (e.g., memory and computational speed), high costs of communication, and limited lifetime. The related literature has recommended different methods, most of which have been based on statistical or similar approaches \cite{Breuniq2000,Saneja2017}. The main objective of such approaches tends to concern approximating the distribution of sensor data,  which in turn can be used to report outliers by computing probabilities or metrics like variance, correlations, mean, etc. \cite{Shahid2015}.\\
Rajasegarar et al. in ~\cite{Rajasegarar2006b} used a cluster-based method, where sensory data were combined into clusters utilizing a static width before using this set-up as the basis of comparison for other sensor nodes. This method did not require any in-depth knowledge of how data was distributed, but it did generate high additional costs in terms of communication.\\
Zhuang and Chen in \cite{Zhuang2006b} proposed two outlier detection techniques. They extract the spatio-temporal correlations of measures that had been detected and attained by several sensor nodes. Rajasegarar et al.'s technique applies a wavelet analysis while Zhuang and Chen's technique uses a method of dynamic time warping. However, both techniques needed to set a specified threshold in order to detect the anomalies.\\
For the detection of outlying sensors and event boundary in SNs, Wu et al. in \cite {Wu2007a} propose two algorithms. The first algorithm starts by calculating, for each sensor, the difference between its reading value and the median reading value obtained from its neighboring reading values. Then, each sensor node collects the differences from its neighborhood and standardizes them. The last step permits the decision of whether the sensor considered is an outlier or not, which is done by comparing the absolute value of its standardized difference with a fixed threshold. If this value is larger than this threshold, the considered sensor is then identified as an outlier. This algorithm is exploited in the second proposed algorithm to localize event sensors at an event boundary. The approach proposed in this paper depends on the specific characteristics/constraints of the communication network and the proposed detection algorithms are based on semi-detected and sometimes incorrect data, which are collected from a randomly selected neighbor.
An enhanced version of the proposed approach in \cite{Wu2007a} is presented in \cite{Guenterberg2007}. In this work, the outlying sensor detection algorithm is enhanced by considering a temporal correlation between sensor nodes. The proposed algorithm in \cite{Guenterberg2007} uses the median of the k nearest neighbors for each sensed data and compare it with the locally saved data in the corresponding sensor. The proposed method improves the accuracy of the detection algorithm but in return, the new proposed algorithm requires additional computational costs.\\
Sheng et al. in ~\cite{Sheng2007} proposed a histogram-based technique that would ensure global outlier detection in WSNs. Rather than sending out all sensory data to the base station, with this method each sensor node kept a summary containing the relevant sensed data on a separate sliding window. Then, using the elaborated summaries collected this way, the base station could extract the distribution of data and filter for typical data only. With this method, outliers tend to be remarked if their measures passed a static threshold value. The principal disadvantage of this work, though, is found in the availability that can occur at unplanned intervals in the base station, and which can cause the shutdown of the entire analysis system. Moreover, this method is limited to applications to one-dimensional data where the spatial distance between the sensor nodes is important.\\
The research conducted by Abid et al. in \cite {Abid2017} proposes a density-based clustering method ordering points for ensuring outlier detection. This method is performed without knowing in advance the number or the labels of the clusters, and it is applied independently of certain constraints related to the considered network (e.g., the topology, the change in scalability, and the form of the collected data). In this work, the "Ordering Points To Identify the Clustering Structure" (OPTICS) method is used to analyze the collected data by applying a density-based clustering algorithm, which ensures the classification of data into events and errors. The limitations of \cite {Abid2017} consist in two major points: 1) the proposed method has a handicap to detect an outlier in a huge number of normal values, and 2) it is more robust to detect possible outliers if the learning window is not very big.\\
Barakkath et al. in \cite{BarakkathNisha2017} proposed a fuzzy-based approach for outlier detection. This work applied a subtractive clustering method. The dataset, which is used for the provided experiments, is divided into multiple sets in which the likenesses within sets are greater than those between the peers. Here, outlier detection is performed by adopting a Takagi-Sugeno fuzzy model to account for the function and selection of parameter membership. In this work, the suggested approach has been applied to a WSN that is divided into clusters and thus is unavailable for application to other networks architectures. In addition, our approach tackles outliers in 2D datasets only, and therefore cannot identify anomalies in datasets with greater dimensions.\\ 
The outlier detection in healthcare applications is studied by Saneja and Rani in \cite {Saneja2017}. In this paper, the authors proposed an approach to outlier detection that was based on the sequential minimalization optimization (SMO) derived from correlation and dynamic regressions. During the initial stage, the values of the correlation coefficient are computed and sorted in order to identify the pairs of strongly correlated sensor nodes. In the second stage, anomalies in individual sensors are identified by applying the sequential minimal optimization regression algorithm (SMOReg). To speed up the processing of big data, the proposed approach in \cite {Saneja2017} relies on a Hadoop MapReduce framework \cite{chebbi2016improvement,chebbi2018comparison}. Despite the high scalability of the proposed approach, the latter is applicable only to data that have linear correlation among the considered attributes, which is not true in certain areas of WSNs where measurements cannot be presented linearly.\\
Identifying outliers may also be performed by calculating the density associated with sensory data measures within a target area. This calculation of density can be executed in an evenly distributed manner. In \cite{Ghalem2019}, a Local Outlier Factor (LOF) method is proposed. This method consists of drawing a circle around "k" measures, where depending on the density level obtained, it attributes an "outlier metric" parameter to each measure, which determines whether or not each such measure should be defined as an outlier. To guarantee a high level of accuracy, it may be necessary to execute the LOF method with numerous values of "k", which in turn may lead to increases in the cost of computation.\\

In \cite{qiao2020fast}, Qiao et al. propose a method combining deep belief network and online quarter-sphere one-class support vector machine to perform outlier detection for large-scale and high-dimensional datasets of WSNs. First, a training process that learns the radius of the quarter sphere is applied. Then, online testing is proposed to perform online outlier detection without supervision. To validate the proposed method, four large-scale datasets having dimensions ranging from 54 to 561 are used. The proposed method is compared with three competitive methods using two metrics, which are classification accuracy and computational time.  In this work, the performance of the proposed method should be demonstrated by its comparison with other outlier detection methods through the computation of additional performance metrics.\\
In \cite{safaeistandalone}, Safaei et al. proposed a local outlier detection algorithm that would run on each individual sensor node of the wireless network under consideration. The proposed approach offered three advantages: 1) a reduction mechanism allowing to eliminate the noneffective features; 2) a prior determination of what size the resulting data histogram memory would be, to ensure efficient use of the available memory; and finally 3) the adaptive Bayesian-network-based classification applied to predict noisy data. Experiments were conducted on real datasets and depicted good accuracy of outlier prediction compared to the existing state-of-the-art methods. This work is applied to ensure only the local outlier detection and the presented experimentation is not extended to include the global outlier detection.\\
Gupta et al. in \cite{gupta2021outlier} employ the Outlierness Factor-based on Neighbourhood (OFN) technique for outlier detection and analysis in sensor networks. In the proposed approach, the neighbourhood points are first determined. Then, the weight of the neighbourhood data is calculated. The OFN technique is employed to classify the outlier data points as events and errors based on spatial and temporal correlations, which are neighbourhood readings and timestamps of readings, respectively. The main disadvantage of the proposed approach is that the experiments presented in this work are conducted using only low dimensional datasets containing between 50 and 100 r- neighbours, which are the nearest neighbours for specific data.\\
A time-series denoising autoencoder (TSDA) network is proposed by Wang et al. in \cite{wang2021ts} to compress the discriminative high-dimensional monitoring data to ensure the representation of the temporal and spatial features of the detection points. In addition, a Gaussian model is used for anomaly point detection in wireless sensor networks. This model is based on auxiliary target variables to gain the anomaly points by employing an objective function of region partitioning. The limitation of the proposed approach is that it performs a slight disadvantage with low-dimensional datasets presenting a limited number of spatial-temporal features.\\
In the next section, we detail our proposed approach for global outlier detection in WSNs.

\section{Proposed Approach for Global Outlier Detection in WSNs}

\begin{figure}
\centering
  \includegraphics[width=\textwidth]{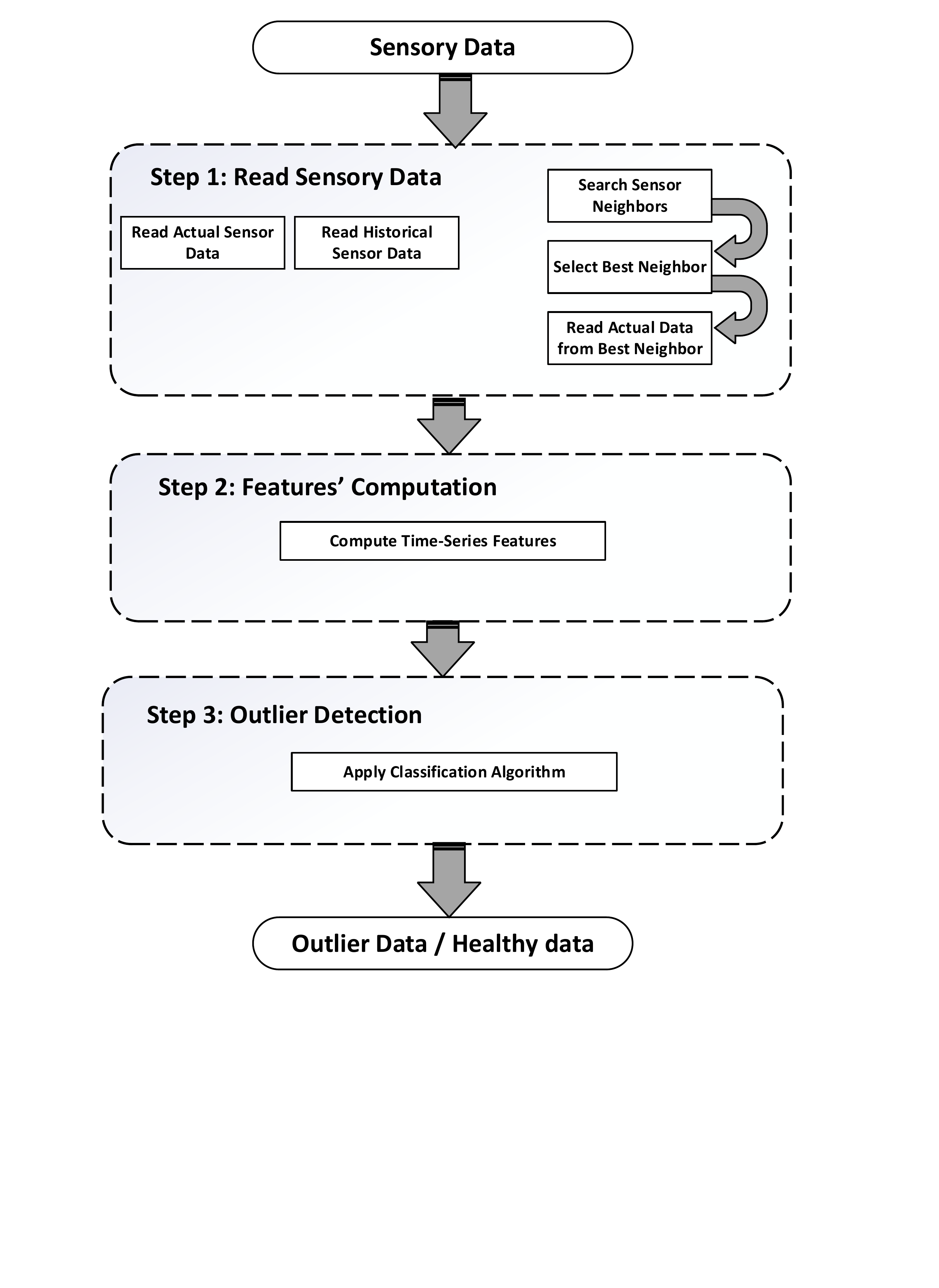}
  \caption{Steps of the Proposed Approach.}
  \label{fig:flow}
\end{figure}

The approach we propose to global outlier detection in WSNs is depicted in Figure \ref{fig:flow}. It is modeled as a process that consists of 3 sequential steps. The first step includes three parallel sub-steps, which are: 1) reading the actual data that are collected from the considered sensor $S$, 2) reading the historical data stored in the memory of $S$, and 3) searching neighbors of $S$, selecting the best neighbor, and reading the actual data from the selected neighbor. The second step aims to ensure the computation of features by using the collected data (i.e., actual data collected from $S$, historical data stored in $S$, and actual data collected from the best neighbor of $S$). The last step applies the outlier detection algorithm to determine outlier data and normal/healthy data.

\subsection{Reading Sensory Data}

This step aims to prepare sensor data to be evaluated for outlier detection. Two types of data are distinguished: data that are specific to a selected sensor and those that are specific to the best neighbor of the considered sensor. Indeed, to detect noise globally, it is necessary to select neighbors that can potentially collaborate with the considered sensor. It is important to determine how many neighbors must be selected and which sensor is the most effective for the collaboration. Hence, for the neighbor selection, a simulation of the Monte Carlo algorithm is conducted. This choice is justified by the fact that this algorithm has shown its usefulness in this context of use, which has been proven in the recently conducted research \cite{Ghalem2019}. After the neighbor selection phase, adaptive entropy and a greedy algorithm are applied to the data that have the same timestamp as their neighbors, and this in order to select the sensor that can collaborate more effectively with the considered sensor to globally detect noise in the considered network.\\
The following subsections outline our process for searching the sensor neighbors and selecting the best one.

\subsubsection{Searching Sensor Neighbors}
To detect the outlier data, every local sensor has to find the best neighbors in order to collaborate with them. This is performed by applying a Monte Carlo simulation. For this purpose, a range of neighbors from $1$ to $10$ has been selected and a matrix $ne$ has been created: $ne = \{1,2,3,4,5,6,7,8,9,10\}$. In addition, $10$ sensors are, randomly selected based on their distance and coverage area and a matrix $D_s$ has been created:  $D_s = \{d_1,d_2,d_3,d_4,d_5,d_6,d_7,d_8,d_9,d_{10}\}$. The result of the simulation shows that the best number of neighbors ensuring an effective spatial collaboration is $4$. The sensor nodes with the nearest distance are more reliable and more accurate compared with others at greater distances. 

\subsubsection{Selecting the Best Neighbor}
After searching neighbors for collaboration, the next step is to calculate and identify the best neighbor. This latter will participate with the local sensor data in the classification algorithm. For this matter, sensors will keep the latest 10 data from the selected neighbors \cite{safaeistandalone}. The input data frame is shown below:
\begin{equation}
D = \left(
\begin{array}{cccccccccc}
  d_{s_1}^{t_n} & d_{s_1}^{t_{n-1}} & d_{s_1}^{t_{n-2}} & d_{s_1}^{t_{n-3}} &d_{s_1}^{t_{n-4}} & d_{s_1}^{t_{n-5}} & d_{s_1}^{t_{n-6}} & d_{s_1}^{t_{n-7}}&d_{s_1}^{t_{n-8}}&d_{s_1}^{t_{n-9}} \\
  d_{s_2}^{t_n} & d_{s_2}^{t_{n-1}} & d_{s_2}^{t_{n-2}} & d_{s_2}^{t_{n-3}} &d_{s_2}^{t_{n-4}} & d_{s_2}^{t_{n-5}} & d_{s_2}^{t_{n-6}} & d_{s_2}^{t_{n-7}}&d_{s_2}^{t_{n-8}}&d_{s_2}^{t_{n-9}}\\
  d_{s_3}^{t_n} & d_{s_3}^{t_{n-1}} & d_{s_3}^{t_{n-2}} & d_{s_3}^{t_{n-3}} &d_{s_3}^{t_{n-4}} & d_{s_3}^{t_{n-5}} & d_{s_3}^{t_{n-6}} & d_{s_3}^{t_{n-7}}&d_{s_3}^{t_{n-8}}&d_{s_3}^{t_{n-9}}\\
  d_{s_4}^{t_n} & d_{s_4}^{t_{n-1}} & d_{s_4}^{t_{n-2}} & d_{s_4}^{t_{n-3}} &d_{s_4}^{t_{n-4}} & d_{s_4}^{t_{n-5}} & d_{s_4}^{t_{n-6}} & d_{s_4}^{t_{n-7}}&d_{s_4}^{t_{n-8}}&d_{s_4}^{t_{n-9}}\\
  \end{array}\right)
\label{eq:dataM}
\end{equation}
The identification of the best neighbor is based on an adaptive entropy function. This function aims to calculate the weight of each sensor node in order to select the best neighbor. This function is deducted from the following equations: \ref{eq:aef}, \ref{eq:eq1}, \ref{eq:eq2}, \ref{eq:eq3}, \ref{eq:eq4}, \ref{eq:eq5}, \ref{eq:eq6}, \ref{eq:eq7}, \ref{eq:eq8}, \ref{eq:eq9}, and \ref{eq:eq10}.
\begin{equation}
d = \{d_{s_x}^{t_n} , d_{s_x}^{t_{n-1}} , d_{s_x}^{t_{n-2}} , d_{s_x}^{t_{n-3}} ,d_{s_x}^{t_{n-4}} , d_{s_x}^{t_{n-5}} , d_{s_x}^{t_{n-6}} , d_{s_x}^{t_{n-7}},d_{s_x}^{t_{n-8}},d_{s_x}^{t_{n-9}}\}
\label{eq:aef}
\end{equation}
Where $d$ is the last $10$ history data of each neighbor sensor ${s_x}$ and ${t_n}$ is the current time. 

\begin{equation}
\overline { x } = \frac {1}{n}  \displaystyle\sum_{i=1}^{10} d_i, \quad n = 10
\label{eq:eq1}
\end{equation}
Where $\overline { x }$ is the mean of the history data of $d$. 
\begin{equation}
e = \frac {d_{s_x}^{t_n} - \overline { x }}{\overline { x }}
\label{eq:eq2}
\end{equation}
$e$ is the deviation of ${d_{s_x}^{t_n}}$ from $\overline { x }$, where  ${t_n}$ is the current time and ${s_x}$ is the neighbor sensor.
\begin{equation}
f_i= \begin{cases}
				h_i = -2 ,& \text{if } e\leq -0.5\\
				h_i = -1 ,& \\text{if } -0.5 < e \leq 0\
				h_i = 1 ,& \text{if } 0 < e \leq 0.5\\
				h_i = 2 ,& \text{if } e > 0.5\\
		\end{cases}
		\label{eq:eq3}
\end{equation}
$h_i$ is the classification of each $e$ value based on the defined condition.
\begin{equation}
a_0 = \displaystyle\sum_{i=1}^{n}{h_i}  \Rightarrow h_i = -2
\label{eq:eq4}
\end{equation}
\begin{equation}
a_1 = \displaystyle\sum_{i=1}^{n}{h_i}  \Rightarrow h_i = -1
\label{eq:eq5}
\end{equation}
\begin{equation}
a_2 = \displaystyle\sum_{i=1}^{n}{h_i}  \Rightarrow h_i = 1
\label{eq:eq6}
\end{equation}
\begin{equation}
a_3 = \displaystyle\sum_{i=1}^{n}{h_i}  \Rightarrow h_i = 2
\label{eq:eq7}
\end{equation}
Where $a_0, a_1, a_2$, and $a_3$ are the total number of ${h_i}$ values.
\begin{equation}
s =  \displaystyle\sum{(a_0, a_1, a_2, a_3)}
\label{eq:eq8}
\end{equation}
$s$ is the sum of all the $(a_0, a_1, a_2, a_3)$ variables.
\begin{equation}
en_t= \begin{cases}
				en_0 = -(\frac{a_0}{s}) \times \log(\frac{a_0}{s})\\
				en_1 = -(\frac{a_1}{s}) \times \log(\frac{a_1}{s})\\
				en_2 = -(\frac{a_2}{s}) \times \log(\frac{a_2}{s})\\
				en_3 = -(\frac{a_3}{s}) \times \log(\frac{a_4}{s})\\
		\end{cases}
		\label{eq:eq9}
\end{equation}
Where $en_t$ is the calculated weight for each variable $a$.
\begin{equation}
\label{eq:eq10}
N_b = max(en_t)
\end{equation}
$N_b$ is the best selected neighbor obtained by choosing the maximum value of $en_t$. 

After selecting the best neighbor, we propose to calculate the corresponding features and build the feature matrix to be used by the classification algorithm. The major problem in WSNs is the limitation of resources such as dependence on batteries as power sources, very limited central processing unit (CPU) and memory capacity, etc. Certainly, increasing the number of features has a direct effect on the outlier detection algorithm's accuracy. However, realistically, it is not feasible to consider multiple features for the case of a single sensor node, and this is due to the previously mentioned limitation of WSNs. In this study, the feature matrix is composed of variables taken from the actual data of the best neighbor (e.g. temperature, pressure, humidity, etc.) and four features computed based on the actual and historical data of the considered sensor.  

\subsection{Features' Computation}

This step is intended to compute a set of features based on collected data (actual data collected from a chosen sensor $S$ and historical data stored in $S$). In this work, four features are computed. These features are Pearson correlation, Spearman ranking correlation, distance correlation, and correlation relationship. These features have been considered in several previous related works and they have provided good results \cite{carvalho2011improving, xie2014segment, jiang2015lifetime, almeida2017improving, li2019anomaly, rajesh2019correlation, angiulli2020reducing}.

\subsubsection{Pearson Correlation Feature}
Examining the relationships between variables is very important in classification algorithms. In this work, we propose to use the "Pearson correlation coefficient", also known as the "product-moment correlation coefficient". This statistical coefficient, denoted in our case by $r$, helps to estimate the relationship between two variables. A value that is close to 0 indicates that there is no relationship between variables, whereas an absolute value that is close to 1 indicates a strong relationship. Generally, the Pearson coefficient is affected by nonlinear behavior. Hence, in our work, the Pearson correlation is measured using an adaptive entropy function to overcome the problem of nonlinear behavior.\\
Let us suppose two variables $x$ and $y$. Equation \ref{eq:PC} demonstrates how the Pearson correlation coefficient between this $x$ and $y$ is calculated:
\begin{equation}
r = \frac { S S _ { x y } } { \sqrt { S S _ { x } S S _ { y } } }
\label{eq:PC}
\end{equation}
$SS_{x}$ and $SS_{y}$ represent the sums of the squared scores of $x$ and $y$, respectively. Whereas, $SS_{xy}$ represent the sum of the products of the squared scores of $x$ and $y$.\\
$SS_{x}$ is calculated using Equation \ref{eq:PCS}, where $\overline {x}$ is the mean of the $x$ sample and $n$ is the size of this sample.
\begin{equation}
S S _ { x } = \sum _ { i = 1 } ^ { n } \left( x _ { i } - \overline { x } \right) ^ { 2 }
\label{eq:PCS}
\end{equation}
The main challenge when calculating $SS_{x}$ using Equation \ref{eq:PCS} is the computing time in case of considering a big dataset. Therefore, the sum of squares can be also calculated using Equation \ref{eq:PC2} in order to overcome the problem of time-consuming computation.
\begin{equation}
S S _ { x } = \sum _ { i = 1 } ^ { n } x _ { i } ^ { 2 } - \frac {\left (\displaystyle \sum _ {i = 1}^{n} x _ { i } \right) ^ { 2 }} { n }
\label{eq:PC2}
\end{equation}
Following the same process, we can calculate the sum of squares for $y$ by modifying $x$ by $y$ in Equation \ref{eq:PC2}.\\
The sum of the products of the squared scores of x and y is computed using Equation \ref{eq:PC4}.

\begin{equation}
S S _ { x y } = \sum _ { i = 1 } ^ { n } \left( x _ { i } y _ { i } \right) - \frac { \left( \displaystyle \sum _ { i = 1 } ^ { n } x _ { i } \right) \left( \displaystyle \sum _ { i = 1 } ^ { n } y _ { i } \right) } { n }
\label{eq:PC4}
\end{equation}

\subsubsection{Spearman  Ranking  Correlation Feature}
The Spearman ranking correlation is a non-parametric coefficient that is used to measure the level of relationship between two variables. This coefficient is suitable for correlation analysis once the variables' values are converted into ordinal scales. Equation \ref{eq:src1} is used to calculate the Spearman ranking correlation coefficient:
\begin{equation}
\rho = 1 - \frac { 6 \sum d _ { i } ^ { 2 } } { n \left( n ^ { 2 } - 1 \right) }
\label{eq:src1}
\end{equation}

The $\rho$ values are between $-1$ and $+1$.\\
When the $\rho$ is close to $-1$ or $+1$, this indicates an important correlation between the considered variables. However, when the value is close to zero, we conclude that there is a weak correlation between the variables.

\subsubsection{Distance Correlation Feature}
To measure the distance correlation between sets of random variables, the Fourier transform is applied.\\
Assume $p$ is a positive number and $X = (X_1,\ldots ,X_p) \in \mathbb {R}^p$ is a random vector. In vector $s = (s_1,\ldots ,s_p) \in \mathbb {R}^p$, the norm $\Vert s\Vert = (s_1^2+\cdots +s_p^2)^{1/2}$ depicts the standard Euclidean norm on $\mathbb {R}^p$.\\
Further, let us consider $\langle s , X \rangle = s _ { 1 } X _ { 1 } + \cdots + s _ { p } X _ { p }$ the standard inner product of $s$ and $X$.\\
Let us also consider the positive numbers $q$ and $a$, a vector $t \in \mathbb {R}^q$, and finally a random vector $Y \in \mathbb {R}^q$. The inner product $\langle t , Y \rangle$ and the Euclidean norm $||t||$ on $\mathbb {R}^q$ are depicted as follows.\\
The common characteristic function of random vectors $(X, Y)$ is given by Equation \ref{eq:ccf}:
\begin{equation}
\phi _ { X , Y } ( s , t ) = \mathbb { E } \exp [ \sqrt { - 1 } \langle s , X \rangle + \sqrt { - 1 } \langle t , Y \rangle ]
\label{eq:ccf}
\end{equation}
Where $\phi _ { X } ( s ) = \phi _ { X , Y } ( s , 0 ) = \mathbb { E } \operatorname { e x p } [ \sqrt { - 1 } \langle s , X \rangle ]$ and $\phi _ { Y } ( t ) = \phi _ { X , Y } ( 0 , t ) = \mathbb { E } \exp [ \sqrt { - 1 } \langle t , Y \rangle ]$ are the marginal characteristic functions of $Y$ and $X$. If $\phi _ { X , Y } ( s , t ) = \phi _ { X } ( s ) \phi Y ( t )$, then 
$X$ and $Y$ are independent for any $s \in \mathbb { R } ^ { p } \text { and } t \in \mathbb { R } ^ { q }$.

For random vectors $X$ and $Y$, the covariance distance is a non-negative number $\mathcal {V}(X,Y)$, here defined by Equation \ref{eq:DC2}:
\begin{equation}
\mathcal { V } ^ { 2 } ( X , Y ) = \frac { 1 } { c _ { p } c _ { q } } \int _ { \mathbb { R } ^ { q } } \int _ { \mathbb { R } ^ { p } } \frac { | \phi _ { X , Y } ( s , t ) - \phi _ { X } ( s ) \phi _ { Y } ( t ) | ^ { 2 } } { \| s \| ^ { p + 1 } \| t \| ^ { q + 1 } } \mathrm { d } s \mathrm { d } t
\label{eq:DC2}
\end{equation}
Where $c _ { p } = \frac { \pi ^ { ( p + 1 ) / 2 } } { \Gamma ( ( p + 1 ) / 2 ) }$.\\
The correlation distance between $X$ and $Y$ is expressed by Equation \ref{eq:DC3}:
\begin{equation}
\mathcal { R } ( X , Y ) = \frac { \mathcal { V } ( X , Y ) } { \sqrt { \mathcal { V } ( X , X ) } \cdot \sqrt { \mathcal { V } ( Y , Y ) } }
\label{eq:DC3}
\end{equation}

The distance correlation is denoted by $T$ and is given by Equation \ref{eq:DC1}. Values of $T$ are in $[0,1]$ and $T$ is equal to zero if $\varphi _ { X , Y } = \varphi _ { X } \varphi _ { Y } \mu - \mathrm { a.e }$.

\begin{equation}
T ( X , Y ; \mu ) = \int _ { \mathbb { R } ^ { p + q } } \left| \varphi _ { X , Y } ( s , t ) - \varphi _ { X } ( s ) \varphi _ { Y } ( t ) \right| ^ { 2 } \mu ( d s , d t )
\label{eq:DC1}
\end{equation}

Where $\varphi _ { X } ( t ) = \mathbb { E } \left[ \mathrm { e } ^ { i \langle t , Z \rangle } \right] , \quad t \in \mathbb { R } ^ { d }$ denotes a characteristic function and $X \in \mathbb { R } ^ { d }$ a random vector.\\
When $\mu$ has a Lebesgue density with positive number on $\mathbb { R } ^ { p + q }$ and if  $T ( X , Y ; \mu ) = 0$, this may result that $X \perp Y$.\\
An empirical version $T_n( X , Y ; \mu )$ of $T( X , Y ; \mu )$ is obtained if attributes in Equation \ref{eq:DC1} are changed by their corresponding empirical versions. Then, based on the distribution of $ T_n $ under the $null$ hypothesis, $X$ and $Y$ are considered as independent.

\subsubsection{Correlation Relationship Feature}
The correlation coefficient, named $r$, allows measuring the linearity relationship between two variables. The correlation coefficient can take any value between $-1$ and $+1$.

The interpretation of the values of the correlation coefficient is as follow:
\begin{itemize}
	\item $0$ demonstrates a non-linear relationship;
	\item $+1$ demonstrates a good "positive linear relationship". When the values of a single variable increase, then the values of another variable will also increase;
	\item $-1$ demonstrates a good "negative linear relationship". When the values of a single variable decrease, then the values of another variable will decrease also;
	\item Values that fall between 0 and 0.3 (or -0.3 and 0) demonstrate a weak positive (negative) relationship using a shaky linear relationship rule;
	\item Values that fall between 0.3 and 0.7 (or -0.7 and -0.3) demonstrate a moderate positive (negative) linear relationship using a fuzzy firm linear rule;
	\item Values that fall between 0.7 and 1.0 (or -1.0 and -0.7) demonstrate a strong positive (negative) linear relationship using a firm linear rule;
	\item The value of $r^2$, also termed the coefficient of determination, shows that $r^2$ tends to be understood as the per cent of the variation of one variable produced by another variable, or the per cent of variation that is shared between two variables. 
\end{itemize}

To calculate the correlation coefficient of two variables $X$ and $Y$, let us consider $zX$ and $zY$ the standardized versions of $X$ and $Y$, respectively. Both $zX$ and $zY$ are re-stated to represent means equaling 0 as well as standard deviations of 1. The expressions we used in order to obtain these standardized scores are represented in equations \ref{eq:COCO1} and \ref{eq:COCO2}:

\begin{equation}
\mathrm { z } \mathrm { X } _ { i } = \left[ \mathrm { X } _ { i } - \operatorname { mean } ( X ) \right] / \mathrm { s.d. } ( X )
\label{eq:COCO1}
\end{equation}
\begin{equation}
\mathrm { z } \mathrm { Y } _ { i } = \left[ \mathrm { Y } _ { i } - \operatorname { mean } ( Y ) \right] / \mathrm { s.d. } ( Y )
\label{eq:COCO2}
\end{equation}

The correlation coefficient can be defined as the mean product of these standardized scores $(zX_i , zY_i )$, as expressed in equation \ref{eq:COCO3}:
\begin{equation}
r _ { X , Y } = \operatorname { sum } \text { of } \left[ \mathrm { zX } _ { i } \times \mathrm { zY } _ { i } \right] / ( n - 1 )
\label{eq:COCO3}
\end{equation}
Where $n$ represents the sample size.

Features that are calculated based on three forms of collected data (actual data from $S$, historical data stored in $S$, and actual data collected from the best neighbor of $S$) are expressed as follow: 

\begin{enumerate}
	\item Pearson Correlation Feature $ \to $ $f_1$
	\item Spearman Rank Correlation Feature$ \to $ $f_2$
	\item Distance Correlation Feature $ \to $ $f_3$
	\item Correlation Coefficient Feature $ \to $ $f_4$
	\item Variable from the actual data collected from the best Neighbor$ \to $ $f_n$ 
\end{enumerate}
Hence, we are able now to construct a feature matrix, denoted by FeatMatrix, that will be used by the outlier detection algorithm as it is shown by Equation \ref{eq:featGO}.
\begin{equation}
FeatMatrix = \left(
\begin{array}{ccccc}
  f_{1_{s_1}} & f_{2_{s_1}} &  f_{3_{s_1}} & f_{4_{s_1}} & f_{n_{s_1}}\\
  f_{1_{s_2}} & f_{2_{s_2}} &  f_{3_{s_2}} & f_{4_{s_2}} & f_{n_{s_2}}\\
  .  & . &  . & . &   .  \\
  .  & . &  . & . &   .  \\
	.  & . &  . & . &   . \\
  f_{1_{s_n}} & f_{2_{s_n}} &  f_{3_{s_n}} & f_{4_{s_n}} & f_{n_{s_n}}\end{array}  \right) 
\label{eq:featGO}
\end{equation}

\subsection{Outlier Detection}
In this work, we compare five different classification algorithms, which are: Random Forest (RF) \cite{breiman2001random}, Naive Bayes (NB) \cite{janakiram2006outlier}, k-Nearest Neighbors (kNN) \cite{sheng2007outlier}, Support Vector Machine (SVM) \cite{zhang2013distributed}, and Neural Network (NN) \cite{yang2009subtractive}. In this work, these five classification algorithms are tested using the features previously detailed in order to determine the best algorithm, which provides the highest accuracy. The proposed experiments are decentralized, with algorithms are running on each sensor node. In this case, it is important to consider the size of the memory that is used by the data history at each node in addition to the accuracy.

\section{Experimentations}
\subsection{Dataset Description}
\begin{figure}
\centering
  \includegraphics[width=0.9\textwidth]{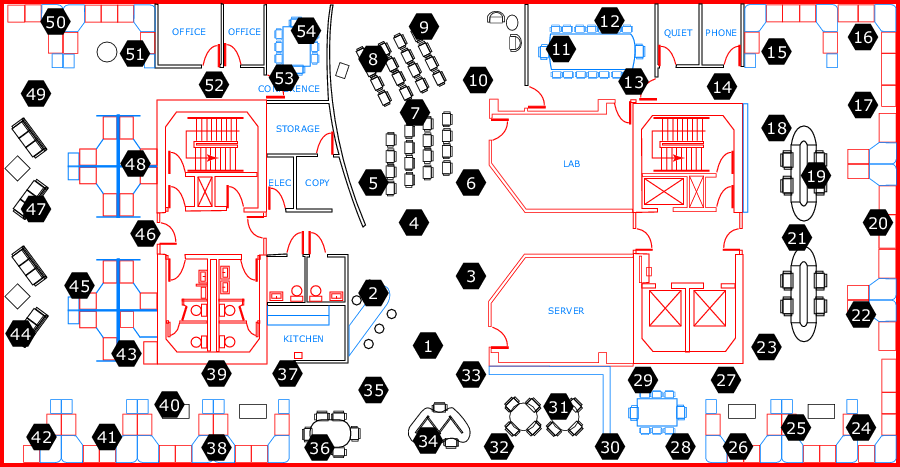}
  \caption{Schematic of the sensors' positioning}
  \label{fig:SensLab}
\end{figure}

This section details the simulation steps followed in order to evaluate the performance of the proposed outlier detection algorithm. MATLAB and R programming tools are used to simulate the results depicted in this study. Experiments were conducted using a dataset from the Intel Berkeley Research lab \cite{Berkeley}, which is one of the most frequently-used datasets in several recent works, such as \cite{Abid2017}. The data collected from $54$ individual sensor nodes deployed in the Intel Berkeley Research Laboratory between February 28$^{th}$ and April 5$^{th}$, 2004 has been gathered in a dataset that includes reading data of approximately $2.3M$ records. In the Intel Berkeley Research Laboratory, Mica2Dot sensors with weatherboards have been used. Mica2Dot sensors are third generation mote modules that are employed to enable the deployment of low power WSNs. These sensors allow the collection of time-stamped topological information, as well as humidity, temperature, light, and voltage values every 31 seconds. These data were collected using the TinyDB network query processing system, built on the TinyOS platform \cite{Berkeley}. Figure \ref{fig:SensLab} presents a schematic of the sensor nodes’ positioning in the test environment thus considered.  

This dataset collects several types of sensory data properties, ranging from temperature, humidity, environmental light, and sensor node battery voltage. Different types of information collected by sensors are displayed in the following formats: "Date ({\ttfamily{yyyy-mm-dd}})," "Time ({\ttfamily{hh:mm:ss.xxx}})," "Epoch ({\ttfamily{Integer}})", "moteid ({\ttfamily{Integer}})," "Temperature ({\ttfamily{Real}}), Humidity ({\ttfamily{Real}})," "Light ({\ttfamily{Real}})," and "Voltage  ({\ttfamily{Real}})".  All data were initially collected on intervals of a $31s$ timestamp.

\subsection{Results and Discussion}
\begin{figure}
  \centering
  \includegraphics[width=1\textwidth]{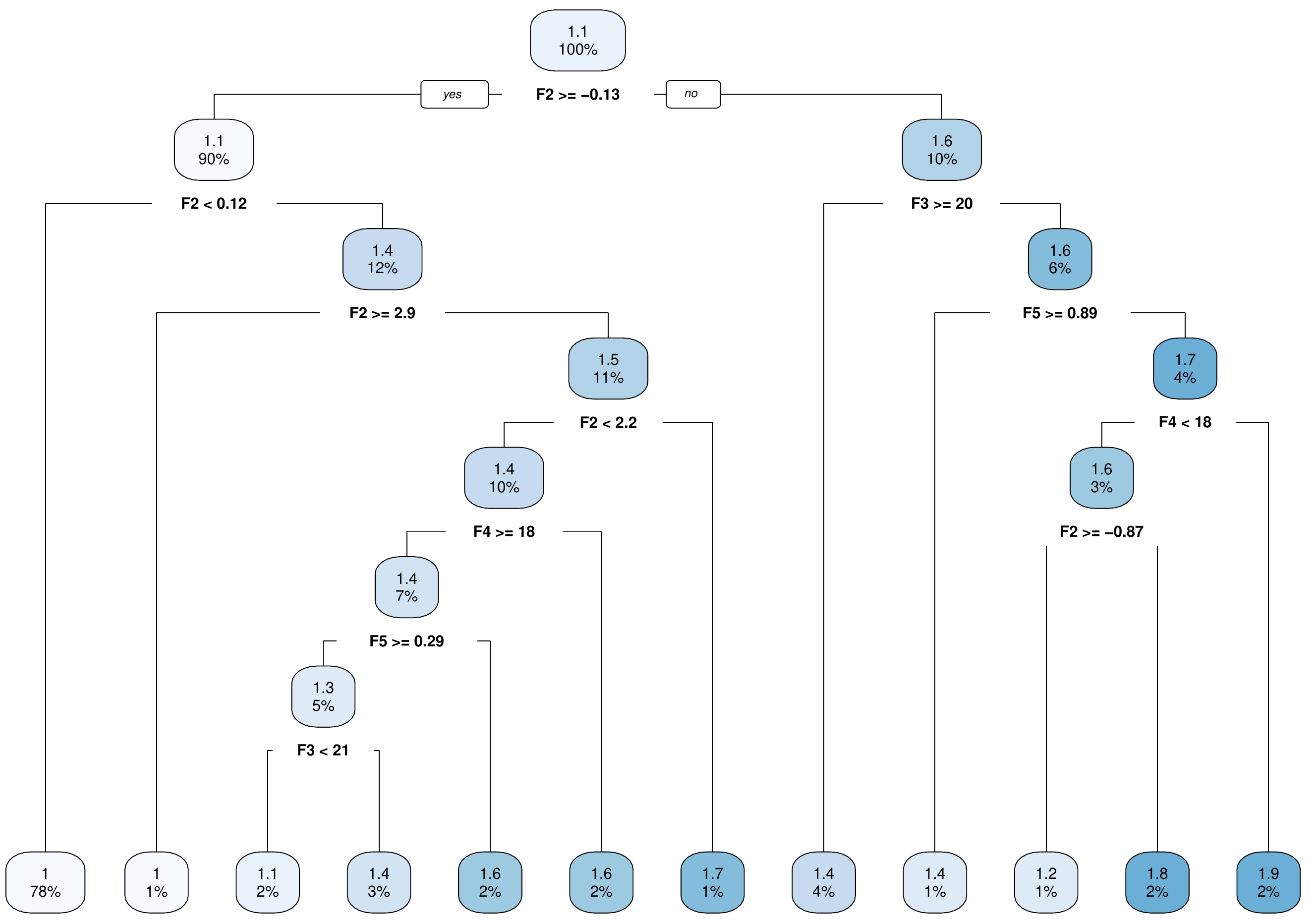}
  \caption{Sample of one decision tree used by the RF algorithm}
  \label{fig:decTreeR}
\end{figure}

To ensure an effective detection of outliers, the best neighbors for each sensor node are selected. Table \ref{tab:NN1} presents sensor neighbors resulting from the execution of the Monte Carlo simulation on the considered scenario, as illustrated by Figure \ref{fig:SensLab}.\\  
\begin{table}
\caption{Sensor nodes with their selected neighbors}
\label{tab:NN1}
\begin{tabular}{cc}
\begin{tabular}{cc}
\hline
Sensor Node&Neighbors\\
\hline\\
S1&S31	,S2	  ,S3	 ,S33  \\
S2&S1	,S3	  ,S4	 ,S35  \\
S3&S1	,S4	  ,S2	 ,S5   \\
S4&S5	,S3	  ,S2	  ,S6  \\
S5&S4	,S6	  ,S3	  ,S9  \\
S6&S9	,S7	  ,S5	  ,S8  \\
S7&S52	,S8	  ,S51	,S6  \\
S8&S7	,S9	  ,S52	,S10 \\
S9&S8	,S10	,S6	 ,S7   \\
S10&S9	,S11	,S12	,S8  \\
S11&S10	,S12	,S13	,S9  \\
S12&S11	,S13	,S10	,S9  \\
S13&S12	,S16	,S11	,S15 \\
S14&S15	,S13	,S16	,S17 \\
S15&S16	,S17	,S14	,S13 \\
S16&S17	,S15	,S13	,S19 \\
S17&S16	,S18	,S19	,S15 \\
S18&S19	,S17	,S20	,S16 \\
S19&S18	,S17	,S20	,S21 \\
S20&S21	,S19	,S18	,S22 \\
S21&S25	,S20	,S19	,S23 \\
S22&S23	,S24	,S20	,S21 \\
S23&S22	,S24	,S25	,S26 \\
S24&S26	,S23	,S25	,S28 \\
S25&S21	,S27	,S24	,S26 \\
S26&S24	,S28	,S25	,S27 \\
\hline
\end{tabular}
&
\begin{tabular}{cc}
\hline
Sensor Node&Neighbors\\
\hline\\
S27&S29	,S25	,S28	,S26 \\
S28&S26	,S29	,S30	,S27 \\
S29&S27	,S28	,S30	,S31 \\
S30&S29	,S28	,S32	,S31 \\
S31&S1	,S29	,S32	,S33 \\
S32&S30	,S33	,S31	,S34 \\
S33&S35	,S32	,S34	,S1  \\
S34&S36	,S33	,S32	,S35 \\
S35&S37	,S33	,S34	,S36 \\
S36&S34	,S38	,S37	,S35 \\
S37&S35	,S38	,S36	,S41 \\
S38&S37	,S39	,S36	,S41 \\
S39&S40	,S38	,S36	,S41 \\
S40&S39	,S38	,S41	,S42 \\
S41&S38	,S37	,S42	,S43 \\
S42&S43	,S41	,S40	,S45 \\
S43&S42	,S44	,S41	,S45 \\
S44&S43	,S45	,S46	,S41 \\
S45&S43	,S44	,S46	,S47 \\
S46&S45	,S47	,S50	,S49 \\
S47&S49	,S48	,S46	,S45 \\
S48&S49	,S47	,S50	,S46 \\
S49&S48	,S47	,S50	,S46 \\
S50&S51	,S49	,S46	,S52 \\
S51&S50	,S52	,S7	  ,S 6 \\
S52&S7	,S51	,S8	  ,S50 \\
\hline
\end{tabular}
\end{tabular}
\end{table}
In this work, five classification algorithms, namely RF, NB, kNN, SVM, and NN, are used to identify outliers in the dataset previously mentioned.\\
The parameters’ values being considered for the five classification algorithms are detailed in our previous work \cite{safaeistandalone}.\\
The number of decision trees is one of the most important parameters in RF algorithms. To obtain an accurate result using the RF method, hundreds or thousands of decision trees are created. When the number of trees increases, the accuracy of results also increases. However, sometimes a larger number of decision trees can affect the system's performance, especially since sensor nodes have limited resources.\\
Figure \ref{fig:decTreeR} shows a sample of a decision tree used by the RF algorithm to ensure outliers' detection. 

\begin{figure}
\centering
  \includegraphics[width=0.9\textwidth]{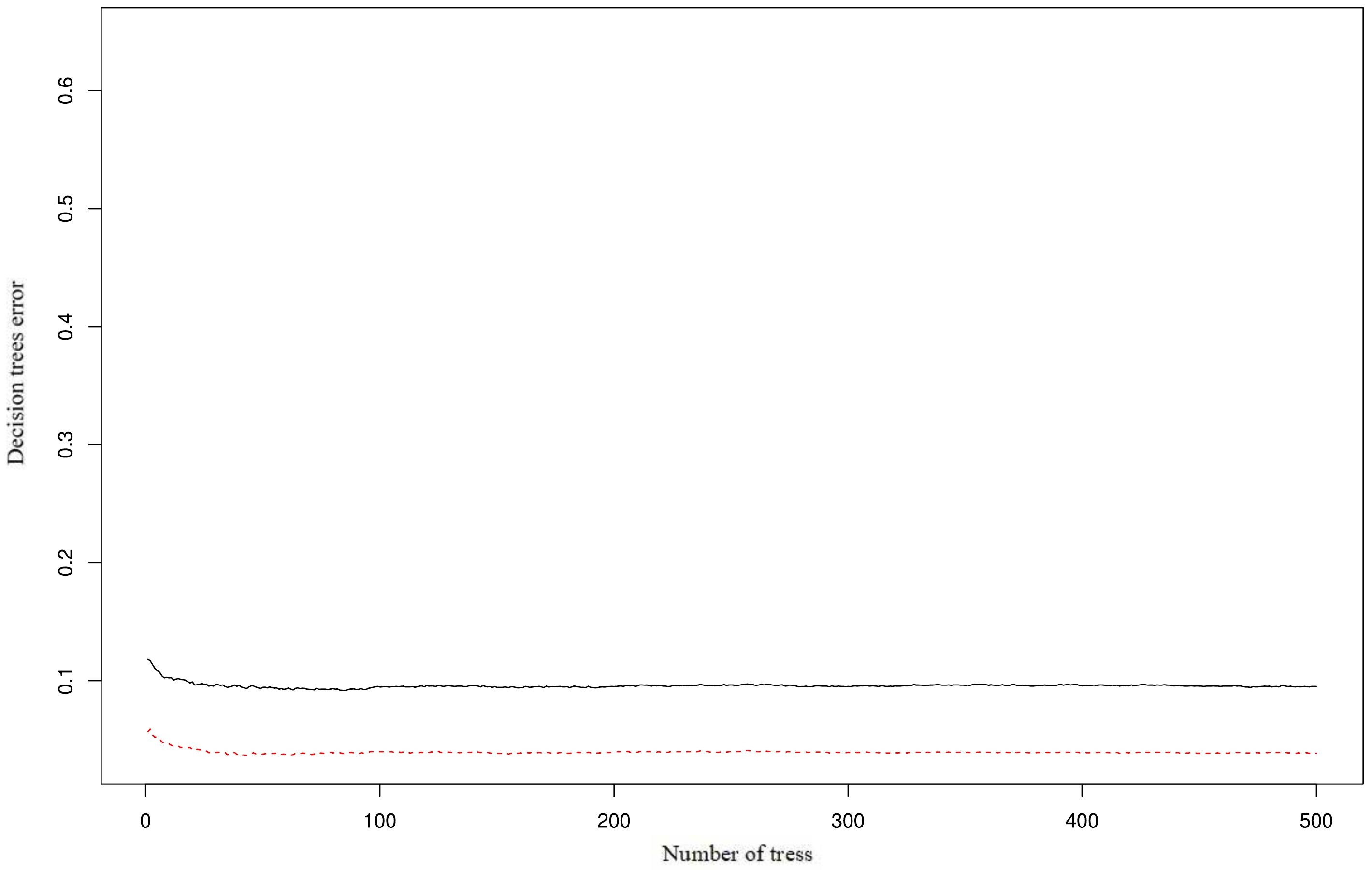}
  \caption{Evaluation of the error rate of the RF algorithm according to the trees' number}
  \label{fig:TreeErr}
\end{figure}

Figure \ref{fig:TreeErr} shows that with $36$ decision tress, RF algorithm achieves the optimum error reduction.  

Table \ref{tab:conf1} illustrates a comparison between the five considered classification algorithms with a noise level of 10\%, 15\%, and 20\% of the total data. "Actual” data represent real data and "prediction” data represent classified data or the output of the classification method. 0 depicts normal data, whereas the value 1 depicts outlier data. The $\Sigma$ represents the sum of values.

\begin{table}
\centering
\caption{Confusion matrices for the five classification algorithms}
\label{tab:conf1}
\setlength{\tabcolsep}{2pt}
\begin{tabular}{p{0.5cm}p{1.5cm}ccc}  
\firsthline
&&\multicolumn{3}{c}{Data with noise level \%} \\
\cline{3-5}
&&10\%& 15\% & 20\%\\
\midrule
\parbox[t]{5mm}{\multirow{20}{*}{\rotatebox[origin=c]{90}{Classification algorithm}}}

&RF & \begin{tabular}{cc|c|c|c}\multicolumn{1}{c}{}&\multicolumn{3}{c}{\tiny{Prediction}}\\&\multicolumn{1}{c}{}&\multicolumn{1}{c}{\tiny{0}}&\multicolumn{1}{c}{\tiny{1}}&\multicolumn{1}{c}{\tiny{$\sum$}} \\\cline{3-4}\parbox[t]{3mm}{\multirow{2}{*}{\rotatebox[origin=c]{90}{\tiny{Actual}}}}& {\tiny{0}}&\cellcolor[gray]{0.5}335724&762&\multicolumn{1}{c}{336486}\\\cline{3-4}&{\tiny{1}}&2236&\cellcolor[gray]{0.5}35158&\multicolumn{1}{c}{37394}\\\cline{3-4}&\multicolumn{1}{c}{\tiny{$\sum$}}&\multicolumn{1}{c}{337960}&\multicolumn{1}{c}{35920}&\multicolumn{1}{c}{373880} \end{tabular}    & \begin{tabular}{cc|c|c|c}\multicolumn{1}{c}{}&\multicolumn{2}{c}{\tiny{Prediction}}\\&\multicolumn{1}{c}{}&\multicolumn{1}{c}{\tiny{0}}&\multicolumn{1}{c}{\tiny{1}}&\multicolumn{1}{c}{\tiny{$\sum$}} \\\cline{3-4}\parbox[t]{3mm}{\multirow{2}{*}{\rotatebox[origin=c]{90}{\tiny{Actual}}}}& {\tiny{0}}&\cellcolor[gray]{0.5}316613&1194&\multicolumn{1}{c}{317807}\\\cline{3-4}&{\tiny{1}}&2721&\cellcolor[gray]{0.5}53352&\multicolumn{1}{c}{56073}\\\cline{3-4}&\multicolumn{1}{c}{\tiny{$\sum$}}&\multicolumn{1}{c}{319334}&\multicolumn{1}{c}{54546}&\multicolumn{1}{c}{373880} \end{tabular}&\begin{tabular}{cc|c|c|c}\multicolumn{1}{c}{}&\multicolumn{2}{c}{\tiny{Prediction}}\\&\multicolumn{1}{c}{}&\multicolumn{1}{c}{\tiny{0}}&\multicolumn{1}{c}{\tiny{1}}&\multicolumn{1}{c}{\tiny{$\sum$}} \\\cline{3-4}\parbox[t]{3mm}{\multirow{2}{*}{\rotatebox[origin=c]{90}{\tiny{Actual}}}}& {\tiny{0}}&\cellcolor[gray]{0.5}297227&1892&\multicolumn{1}{c}{299119}\\\cline{3-4}&{\tiny{1}}&3147&\cellcolor[gray]{0.5}71614&\multicolumn{1}{c}{74761}\\\cline{3-4}&\multicolumn{1}{c}{\tiny{$\sum$}}&\multicolumn{1}{c}{300374}&\multicolumn{1}{c}{73506}&\multicolumn{1}{c}{373880} \end{tabular}    \\
\cline{2-5}
&kNN& \begin{tabular}{cc|c|c|c}\multicolumn{1}{c}{}&\multicolumn{3}{c}{\tiny{Prediction}}\\&\multicolumn{1}{c}{}&\multicolumn{1}{c}{\tiny{0}}&\multicolumn{1}{c}{\tiny{1}}&\multicolumn{1}{c}{\tiny{$\sum$}} \\\cline{3-4}\parbox[t]{3mm}{\multirow{2}{*}{\rotatebox[origin=c]{90}{\tiny{Actual}}}}& {\tiny{0}}&\cellcolor[gray]{0.5}332159&4327&\multicolumn{1}{c}{336486}\\\cline{3-4}&{\tiny{1}}&10165&\cellcolor[gray]{0.5}27229&\multicolumn{1}{c}{37394}\\\cline{3-4}&\multicolumn{1}{c}{\tiny{$\sum$}}&\multicolumn{1}{c}{342324}&\multicolumn{1}{c}{31556}&\multicolumn{1}{c}{373880} \end{tabular} & \begin{tabular}{cc|c|c|c}\multicolumn{1}{c}{}&\multicolumn{2}{c}{\tiny{Prediction}}\\&\multicolumn{1}{c}{}&\multicolumn{1}{c}{\tiny{0}}&\multicolumn{1}{c}{\tiny{1}}&\multicolumn{1}{c}{\tiny{$\sum$}} \\\cline{3-4}\parbox[t]{3mm}{\multirow{2}{*}{\rotatebox[origin=c]{90}{\tiny{Actual}}}}& {\tiny{0}}&\cellcolor[gray]{0.5}309955&7852&\multicolumn{1}{c}{317807}\\\cline{3-4}&{\tiny{1}}&15325&\cellcolor[gray]{0.5}40748&\multicolumn{1}{c}{56073}\\\cline{3-4}&\multicolumn{1}{c}{\tiny{$\sum$}}&\multicolumn{1}{c}{325280}&\multicolumn{1}{c}{48600}&\multicolumn{1}{c}{373880} \end{tabular}&\begin{tabular}{cc|c|c|c}\multicolumn{1}{c}{}&\multicolumn{2}{c}{\tiny{Prediction}}\\&\multicolumn{1}{c}{}&\multicolumn{1}{c}{\tiny{0}}&\multicolumn{1}{c}{\tiny{1}}&\multicolumn{1}{c}{\tiny{$\sum$}} \\\cline{3-4}\parbox[t]{3mm}{\multirow{2}{*}{\rotatebox[origin=c]{90}{\tiny{Actual}}}}& {\tiny{0}}&\cellcolor[gray]{0.5}287298&11821&\multicolumn{1}{c}{299119}\\\cline{3-4}&{\tiny{1}}&20666&\cellcolor[gray]{0.5}54095&\multicolumn{1}{c}{74761}\\\cline{3-4}&\multicolumn{1}{c}{\tiny{$\sum$}}&\multicolumn{1}{c}{307964}&\multicolumn{1}{c}{65916}&\multicolumn{1}{c}{373880} \end{tabular} \\
\cline{2-5}
&NB& \begin{tabular}{cc|c|c|c}\multicolumn{1}{c}{}&\multicolumn{3}{c}{\tiny{Prediction}}\\&\multicolumn{1}{c}{}&\multicolumn{1}{c}{\tiny{0}}&\multicolumn{1}{c}{\tiny{1}}&\multicolumn{1}{c}{\tiny{$\sum$}} \\\cline{3-4}\parbox[t]{3mm}{\multirow{2}{*}{\rotatebox[origin=c]{90}{\tiny{Actual}}}}& {\tiny{0}}&\cellcolor[gray]{0.5}332089&4397&\multicolumn{1}{c}{336486}\\\cline{3-4}&{\tiny{1}}&10000&\cellcolor[gray]{0.5}27394&\multicolumn{1}{c}{37394}\\\cline{3-4}&\multicolumn{1}{c}{\tiny{$\sum$}}&\multicolumn{1}{c}{342089}&\multicolumn{1}{c}{31791}&\multicolumn{1}{c}{373880} \end{tabular}     & \begin{tabular}{cc|c|c|c}\multicolumn{1}{c}{}&\multicolumn{2}{c}{\tiny{Prediction}}\\&\multicolumn{1}{c}{}&\multicolumn{1}{c}{\tiny{0}}&\multicolumn{1}{c}{\tiny{1}}&\multicolumn{1}{c}{\tiny{$\sum$}} \\\cline{3-4}\parbox[t]{3mm}{\multirow{2}{*}{\rotatebox[origin=c]{90}{\tiny{Actual}}}}& {\tiny{0}}&\cellcolor[gray]{0.5}299812&17995&\multicolumn{1}{c}{317807}\\\cline{3-4}&{\tiny{1}}&45810&\cellcolor[gray]{0.5}10263&\multicolumn{1}{c}{56073}\\\cline{3-4}&\multicolumn{1}{c}{\tiny{$\sum$}}&\multicolumn{1}{c}{345622}&\multicolumn{1}{c}{28258}&\multicolumn{1}{c}{373880} \end{tabular}&\begin{tabular}{cc|c|c|c}\multicolumn{1}{c}{}&\multicolumn{2}{c}{\tiny{Prediction}}\\&\multicolumn{1}{c}{}&\multicolumn{1}{c}{\tiny{0}}&\multicolumn{1}{c}{\tiny{1}}&\multicolumn{1}{c}{\tiny{$\sum$}} \\\cline{3-4}\parbox[t]{3mm}{\multirow{2}{*}{\rotatebox[origin=c]{90}{\tiny{Actual}}}}& {\tiny{0}}&\cellcolor[gray]{0.5}260830&38289&\multicolumn{1}{c}{299119}\\\cline{3-4}&{\tiny{1}}&48251&\cellcolor[gray]{0.5}26510&\multicolumn{1}{c}{74761}\\\cline{3-4}&\multicolumn{1}{c}{\tiny{$\sum$}}&\multicolumn{1}{c}{309081}&\multicolumn{1}{c}{64799}&\multicolumn{1}{c}{373880} \end{tabular}\\
\cline{2-5}
&SVM& \begin{tabular}{cc|c|c|c}\multicolumn{1}{c}{}&\multicolumn{3}{c}{\tiny{Prediction}}\\&\multicolumn{1}{c}{}&\multicolumn{1}{c}{\tiny{0}}&\multicolumn{1}{c}{\tiny{1}}&\multicolumn{1}{c}{\tiny{$\sum$}} \\\cline{3-4}\parbox[t]{3mm}{\multirow{2}{*}{\rotatebox[origin=c]{90}{\tiny{Actual}}}}& {\tiny{0}}&\cellcolor[gray]{0.5}171216&165270&\multicolumn{1}{c}{336486}\\\cline{3-4}&{\tiny{1}}&22640&\cellcolor[gray]{0.5}14754&\multicolumn{1}{c}{37394}\\\cline{3-4}&\multicolumn{1}{c}{\tiny{$\sum$}}&\multicolumn{1}{c}{193856}&\multicolumn{1}{c}{180024}&\multicolumn{1}{c}{373880} \end{tabular}& \begin{tabular}{cc|c|c|c}\multicolumn{1}{c}{}&\multicolumn{3}{c}{\tiny{Prediction}}\\&\multicolumn{1}{c}{}&\multicolumn{1}{c}{\tiny{0}}&\multicolumn{1}{c}{\tiny{1}}&\multicolumn{1}{c}{\tiny{$\sum$}} \\\cline{3-4}\parbox[t]{3mm}{\multirow{2}{*}{\rotatebox[origin=c]{90}{\tiny{Actual}}}}& {\tiny{0}}&\cellcolor[gray]{0.5}159517&158290&\multicolumn{1}{c}{317807}\\\cline{3-4}&{\tiny{1}}&33270&\cellcolor[gray]{0.5}22803&\multicolumn{1}{c}{56073}\\\cline{3-4}&\multicolumn{1}{c}{\tiny{$\sum$}}&\multicolumn{1}{c}{192787}&\multicolumn{1}{c}{181093}&\multicolumn{1}{c}{373880} \end{tabular} &\begin{tabular}{cc|c|c|c}\multicolumn{1}{c}{}&\multicolumn{2}{c}{\tiny{Prediction}}\\&\multicolumn{1}{c}{}&\multicolumn{1}{c}{\tiny{0}}&\multicolumn{1}{c}{\tiny{1}}&\multicolumn{1}{c}{\tiny{$\sum$}} \\\cline{3-4}\parbox[t]{3mm}{\multirow{2}{*}{\rotatebox[origin=c]{90}{\tiny{Actual}}}}& {\tiny{0}}&\cellcolor[gray]{0.5}144626&154493&\multicolumn{1}{c}{299119}\\\cline{3-4}&{\tiny{1}}&44063&\cellcolor[gray]{0.5}30698&\multicolumn{1}{c}{74761}\\\cline{3-4}&\multicolumn{1}{c}{\tiny{$\sum$}}&\multicolumn{1}{c}{188689}&\multicolumn{1}{c}{185191}&\multicolumn{1}{c}{373880} \end{tabular}\\
\cline{2-5}
&NN& \begin{tabular}{cc|c|c|c}\multicolumn{1}{c}{}&\multicolumn{3}{c}{\tiny{Prediction}}\\&\multicolumn{1}{c}{}&\multicolumn{1}{c}{\tiny{0}}&\multicolumn{1}{c}{\tiny{1}}&\multicolumn{1}{c}{\tiny{$\sum$}} \\\cline{3-4}\parbox[t]{3mm}{\multirow{2}{*}{\rotatebox[origin=c]{90}{\tiny{Actual}}}}& {\tiny{0}}&\cellcolor[gray]{0.5}171216&165270&\multicolumn{1}{c}{336486}\\\cline{3-4}&{\tiny{1}}&22640&\cellcolor[gray]{0.5}1475&\multicolumn{1}{c}{37394}\\\cline{3-4}&\multicolumn{1}{c}{\tiny{$\sum$}}&\multicolumn{1}{c}{193856}&\multicolumn{1}{c}{180024}&\multicolumn{1}{c}{373880} \end{tabular}     & \begin{tabular}{cc|c|c|c}\multicolumn{1}{c}{}&\multicolumn{2}{c}{\tiny{Prediction}}\\&\multicolumn{1}{c}{}&\multicolumn{1}{c}{\tiny{0}}&\multicolumn{1}{c}{\tiny{1}}&\multicolumn{1}{c}{\tiny{$\sum$}} \\\cline{3-4}\parbox[t]{3mm}{\multirow{2}{*}{\rotatebox[origin=c]{90}{\tiny{Actual}}}}& {\tiny{0}}&\cellcolor[gray]{0.5}312303&5504&\multicolumn{1}{c}{317807}\\\cline{3-4}&{\tiny{1}}&25386&\cellcolor[gray]{0.5}30687&\multicolumn{1}{c}{56073}\\\cline{3-4}&\multicolumn{1}{c}{\tiny{$\sum$}}&\multicolumn{1}{c}{337689}&\multicolumn{1}{c}{36191}&\multicolumn{1}{c}{373880} \end{tabular} &\begin{tabular}{cc|c|c|c}\multicolumn{1}{c}{}&\multicolumn{2}{c}{\tiny{Prediction}}\\&\multicolumn{1}{c}{}&\multicolumn{1}{c}{\tiny{0}}&\multicolumn{1}{c}{\tiny{1}}&\multicolumn{1}{c}{\tiny{$\sum$}} \\\cline{3-4}\parbox[t]{3mm}{\multirow{2}{*}{\rotatebox[origin=c]{90}{\tiny{Actual}}}}& {\tiny{0}}&\cellcolor[gray]{0.5}277344&21775&\multicolumn{1}{c}{299119}\\\cline{3-4}&{\tiny{1}}&20156&\cellcolor[gray]{0.5}54605&\multicolumn{1}{c}{74761}\\\cline{3-4}&\multicolumn{1}{c}{\tiny{$\sum$}}&\multicolumn{1}{c}{297500}&\multicolumn{1}{c}{76380}&\multicolumn{1}{c}{373880} \end{tabular}     \\
\bottomrule
\end{tabular}
\end{table}

\begin{figure}
  \centering
  \includegraphics[width=1\textwidth]{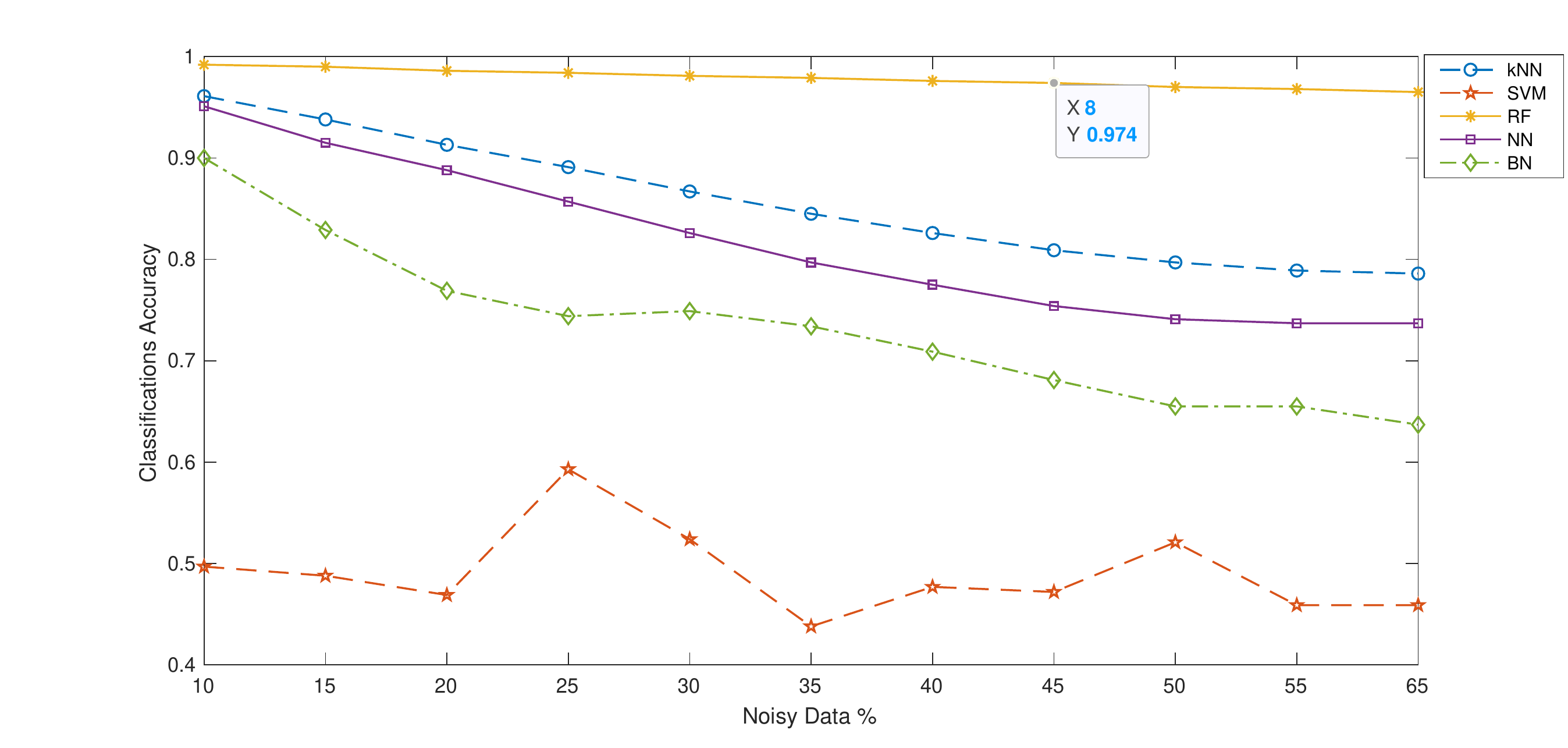}
  \caption{Comparison of the accuracy between the five classification algorithms according to the percentage of noisy data}
  \label{fig:WSNApp}
\end{figure}

Figure \ref{fig:WSNApp} depicts a comparison of the accuracy of the outlier detection between the five considered classification algorithms. This figure shows that an RF algorithm can detect the outlier data with 99.1\% accuracy in 10\% noisy sensory data followed by kNN, NN, NB, and SVM. The outlier detection accuracy of RF will decrease very slowly with the increase of the noisy data but it still has the best accuracy compared to the other algorithms. With a huge amount of noisy sensory data, RF can detect 97.8\% of outlier data but the accuracy of the kNN algorithm drops dramatically to less than 80\%. In this study, SVM provides the most inaccurate results for the outlier detection problem compared to the other four classification algorithms. An important extension of this work will be to combine results of the five classifiers instead of using only one of them \cite{boulila2009improving,farouq2019novel}.

\begin{figure}
  \centering
  \includegraphics[width=\textwidth]{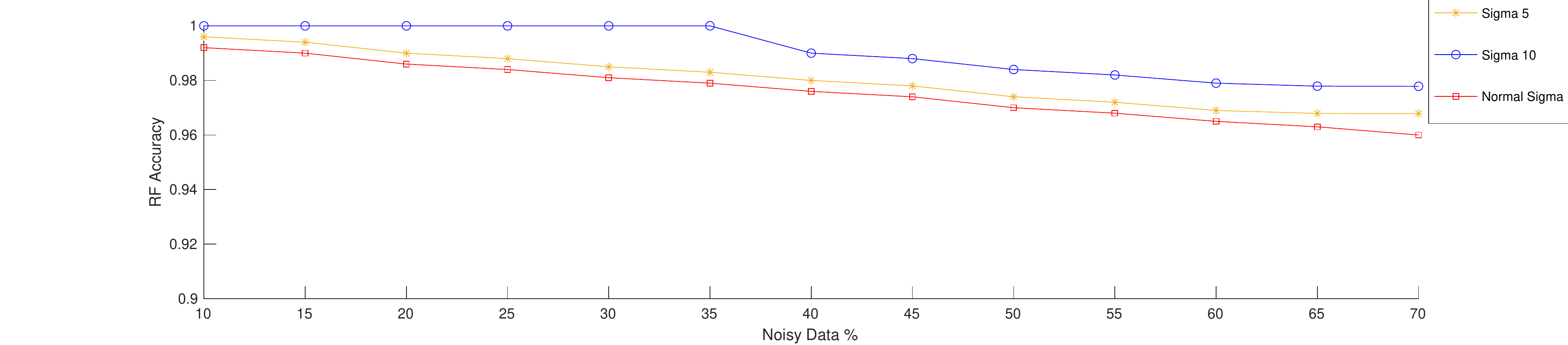}
  \caption{Accuracy of the RF outlier detection algorithm in a noisy environment when $\sigma$ = 0, 5 and 10}
  \label{fig:ADRFS5}
\end{figure}

When the level of noise increases, the gap between noisy data and healthy data will also increase; therefore, the classification algorithms can detect outliers more accurately. Figure \ref{fig:ADRFS5} shows a scenario with a noise-level $\sigma$ equals to 0, 5, and 10. For this scenario, the accuracy of the RF algorithm has increased from 98\% to 99\%. 

\begin{figure}
  \centering
  \includegraphics[width=\textwidth]{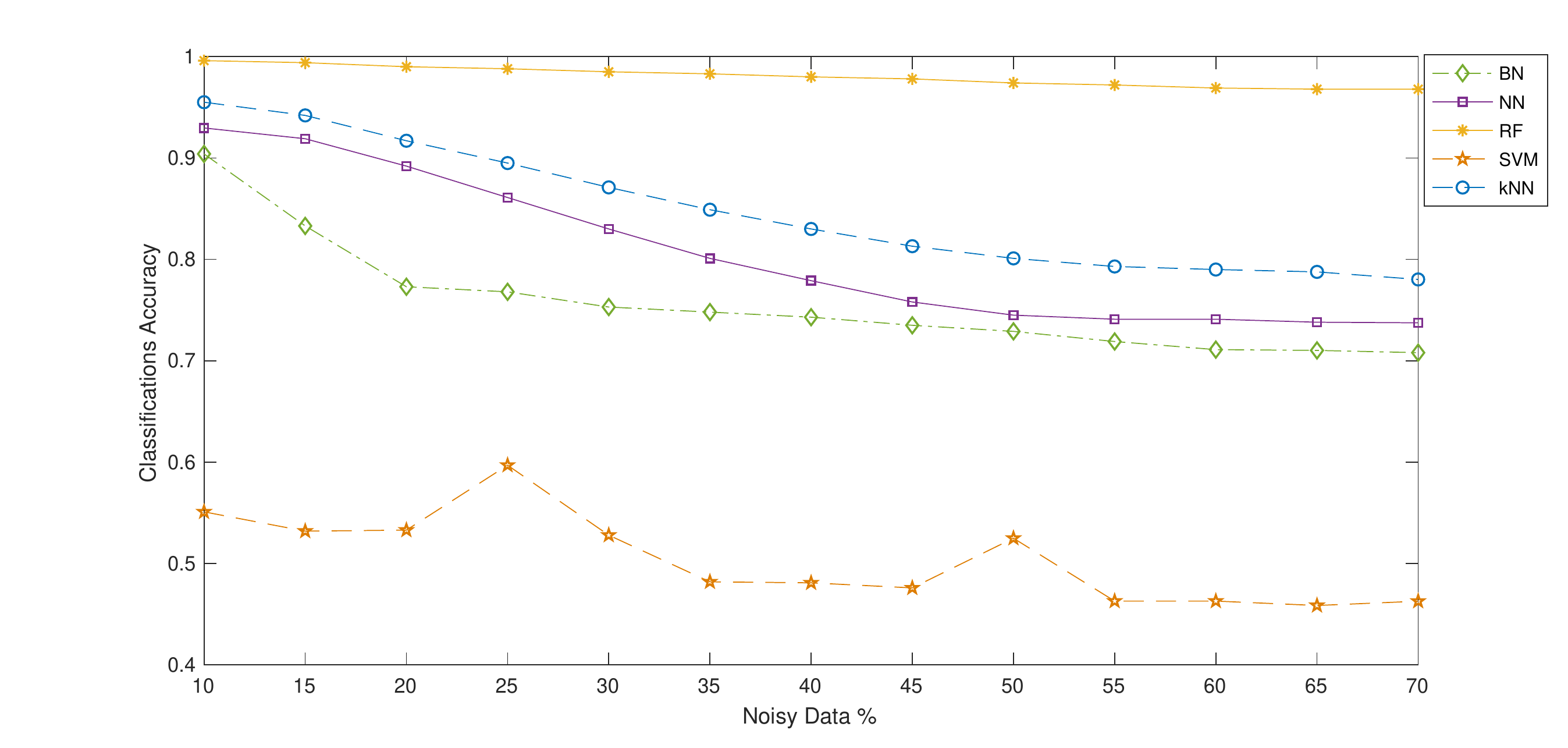}
  \caption{Accuracy of the five classification algorithms in a noisy environment when $\sigma = 5$}
  \label{fig:ADRFS51}
\end{figure}

\begin{figure}
  \centering
  \includegraphics[width=\textwidth]{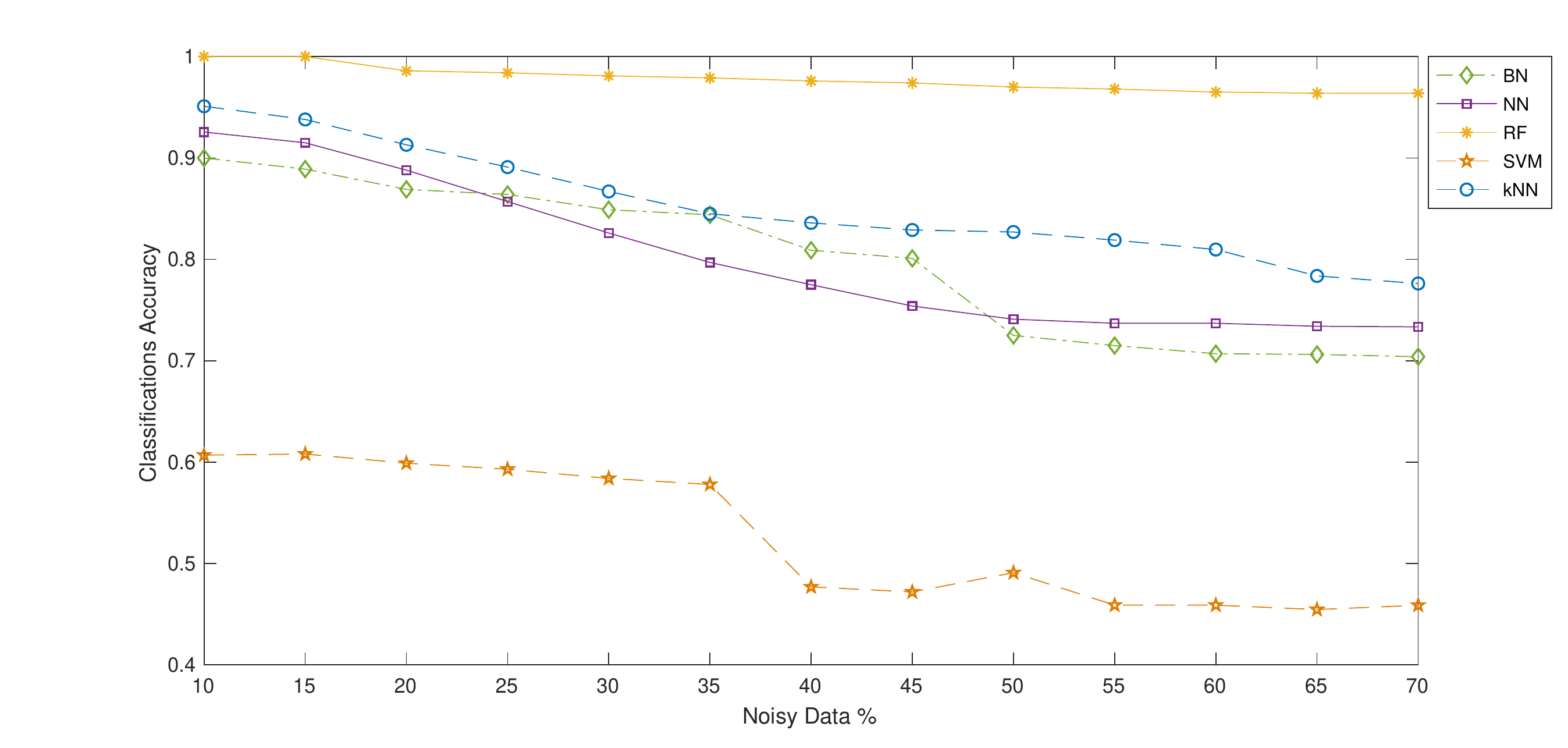}
  \caption{Accuracy of the five classification algorithms in a noisy environment when $\sigma = 7.5$}
  \label{fig:ADRFS75}
\end{figure}

\begin{figure}
  \centering
  \includegraphics[width=\textwidth]{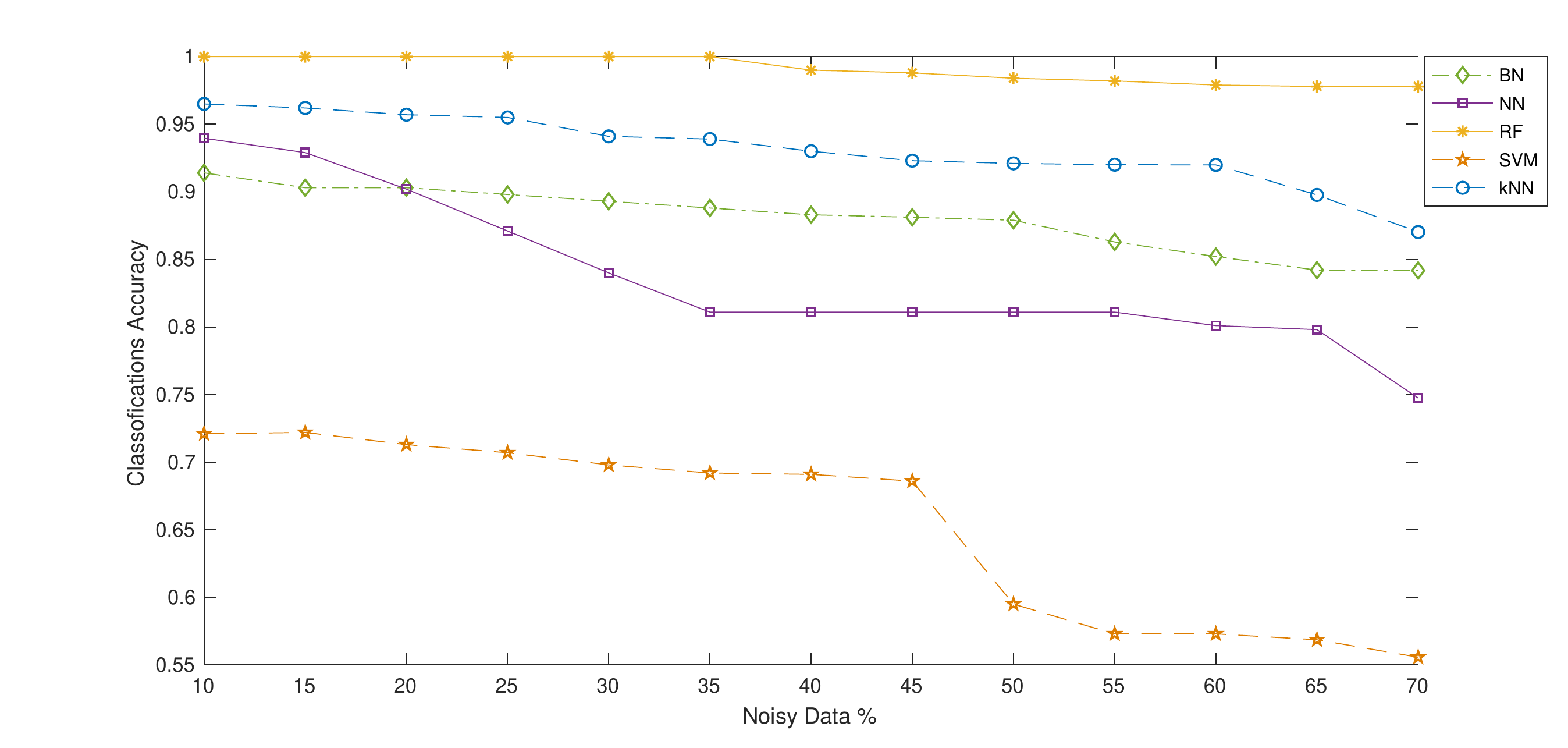}
  \caption{Accuracy of the five classification algorithms in a noisy environment when $\sigma = 10$}
  \label{fig:ADRFS10}
\end{figure}

But $\sigma$, or noise level, is not the only factor influencing the outlier detection accuracy. Another important factor is the total of noisy data in the considered dataset. This work shows that when the amount of noisy data increases, the accuracy of the outlier detection algorithm decreases. Figure \ref{fig:ADRFS5} demonstrates that when the $\sigma$ value increases, the algorithm can identify 100\% of outliers data up to 20\% of noisy data.\\
In this work, three simulation rounds were executed with the same configuration and while changing the value of $\sigma$ in each round to test the accuracy and the behavior of the five classification algorithms (RF, NB, kNN, SVM, and NN). The value of $\sigma$ was changed during the simulation rounds as follow $\sigma = \{5, 7.5, 10\}$. The increase of the value of $\sigma$ has a direct effect on the increase of the noise level since $\sigma$ is one of the main values in the Gaussian noise. Figures \ref{fig:ADRFS51}, \ref{fig:ADRFS75}, and \ref{fig:ADRFS10} show that the accuracy of the outlier detection algorithms changes when the $\sigma$ value increases. Due to the increase of the noise level in the dataset, most of the algorithms can classify the outlier data from normal data more accurately. However, for the SVM algorithm and with increasing the $\sigma$ value, the behavior of outlier detection has been changed from a random prediction to a flow of prediction and classification (the SVM graph shape becomes smoother). This shows that the SVM cannot classify the outlier data with small amounts of noise, but regardless, this does not mean that the classification accuracy provided by SVM has gradually changed.\\
The RF accuracy increased and reached the maximum classification accuracy, which is $100\%$ at certain points of the simulation rounds. The accuracy percentage increased rapidly when the value of $\sigma$ increased from $7.5$ to $10$. Indeed, the overall accuracy, when $\sigma$ is equal to $10$, is more than $99.7\%$ for all percentages of noisy data.\\
For kNN and NB algorithms, the accuracy of outlier detection has been increased with the increase of the value of $\sigma$. The accuracy of NB, in particular, has increased very quickly compared to kNN, but overall, as shown in Figure \ref{fig:ADRFS10}, the accuracy of kNN is higher than that of NB. Concerning NN, in some parts of the simulation, it shows the same value and this is due to the problems of fitting or stack at epoch. But overall, the output analysis shows that the RF algorithm provides the highest outlier detection accuracy compared to the other classification algorithms.

\begin{figure}
  \centering
  \includegraphics[width=1\textwidth]{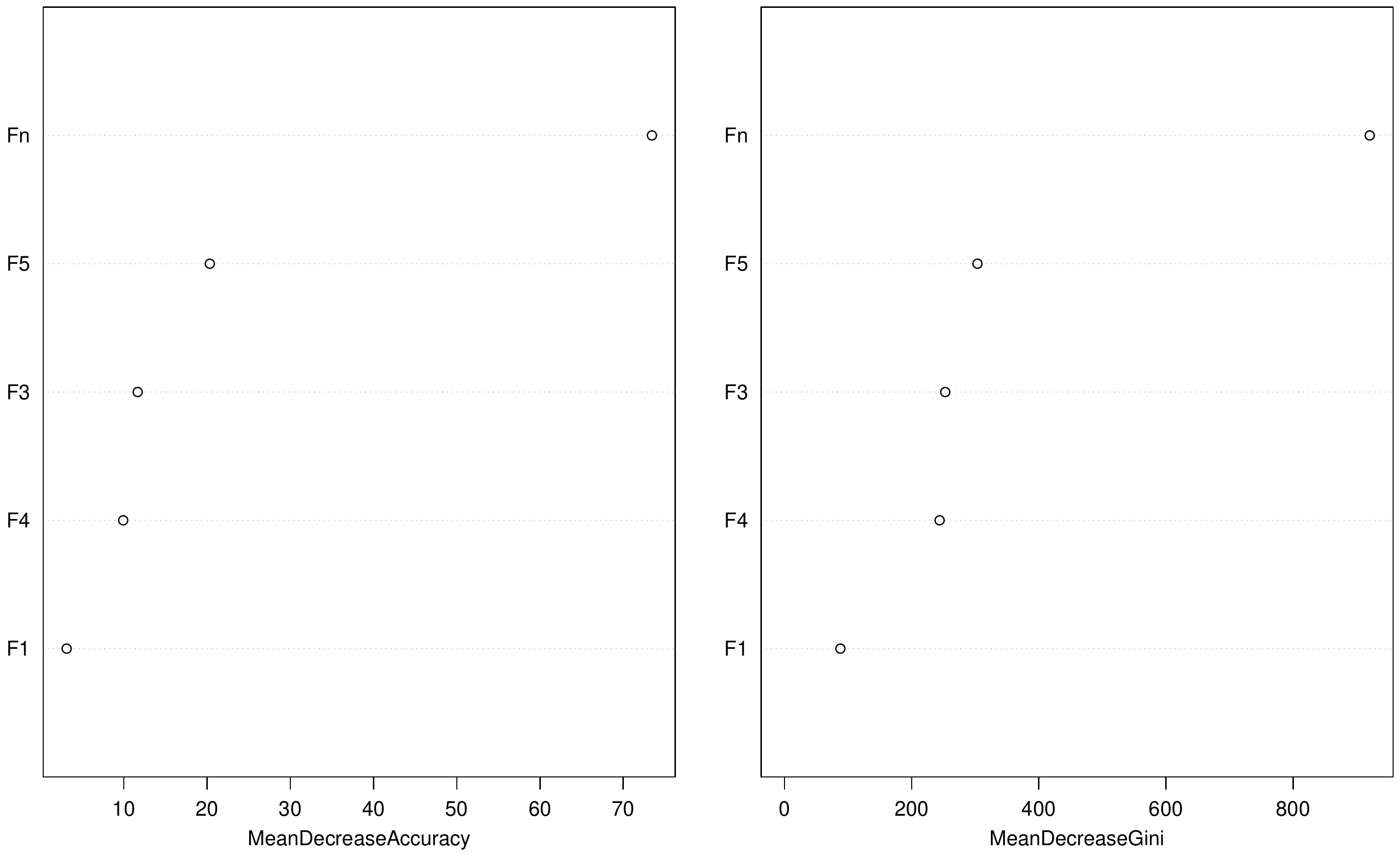}
  \caption[Importance variable comparison]{Comparison of features' importance for the global outlier detection algorithm}
  \label{fig:VarImpot}
\end{figure}

Finally, identifying the importance of features that have been considered for the RF algorithm requires discussion in this section. Figure \ref{fig:VarImpot} shows the importance of the features based on two measurements: Mean Decrease Accuracy (MDA) and Mean Decrease in Gini (MDG). The first one shows how much the accuracy will be reduced if we exclude each feature from the proposed algorithm. For instance, Fn has the highest impact on the algorithm accuracy, which means that without this feature the proposed algorithm can detect outliers but probably with a very low accuracy reaching less than 30\%. The considered features are plotted in descending importance; a feature with a high accuracy means that considering this feature will lead to better outlier detection. The second measurement, MDG, depicts the impurity of features. It is used as a metric to divide data into smaller groups in the decision tree; therefore, the MDG shows how pure the nodes are at the end of the tree.

\section{Conclusion}
This paper proposes a novel global outlier detection approach for WSNs. Our approach is based on time-series analysis, entropy technique, and random forest-based classification algorithm. This approach allows for the utilization of actual sensory data, as well as historical data and data collected from the best neighbor, in order to detect outliers. Experimental results obtained from a real and synthetic dataset have proven the capabilities of our proposed detection approach to adapt its behavior to suit different dynamics and noise level scenarios, thus achieving a significant classification accuracy compared to existing non-time-series approaches. In future work, this approach can be enhanced by proposing effective solutions for the sensor nodes’ detention problem in WSNs, which can prevent further negative effects on the decision-making process. In addition, we plan to consider different datasets to conduct more comprehensive experiments allowing to confirm the effectiveness of the proposed approach. Finally, an interesting perspective of the present work
would be to investigate the impact of varying the number of features on the proposed approach performance.

\clearpage

\bibliography{ElsPaper,Outlier,OutlierWSN,Generals}

\end{document}